%% file: 0-main.tex
\documentclass[conference]{IEEEtran}
\IEEEoverridecommandlockouts

\makeatletter
\def\endthebibliography{%
  \def\@noitemerr{\@latex@warning{Empty `thebibliography' environment}}%
  \endlist
}

\usepackage[]{inputenc}
\usepackage[T1]{fontenc}

\usepackage[numbers]{natbib}
\usepackage{listings}
\usepackage{multirow}

\usepackage{booktabs} % For formal tables
\usepackage{mdframed} 
\usepackage{framed}
\usepackage[x11names]{xcolor}
\colorlet{shadecolor}{blue!20}
\usepackage{makecell}

\usepackage{amsmath,amssymb,amsfonts}
\usepackage{algorithmic}
\usepackage{graphicx}
\usepackage{textcomp}
\usepackage{xcolor}
\usepackage{tabularx}
\usepackage{comment}

\usepackage{rotating}
\usepackage{tikz}

\begin{document}

\title{How Do Code Changes Evolve in Different Platforms? A Mining-based Investigation\\
}
%\author{Authors were omitted due to double-blind review process}

% \and
% \and
% \and
% \and

\author{\IEEEauthorblockN{Markos Viggiato$^1$, Johnatan Oliveira$^2$, Eduardo Figueiredo$^2$, Pooyan Jamshidi$^3$, Christian K\"{a}stner$^4$}
\IEEEauthorblockA{\textit{$^1$Dept. of Electrical and Computer Engineering, University of Alberta}, Edmonton, Canada\\ \textit{$^2$Computer Science Department}, \textit{Federal University of Minas Gerais},  Belo Horizonte, Brazil\\  \textit{$^3$Computer Science and Engineering Department}, \textit{University of South Carolina}, Columbia, United States\\
\textit{$^4$Institute for Software Research}, \textit{ Carnegie Mellon University},
Pittsburgh, United States}} 
%viggiato@ualberta.ca, \{johnatan.si, figueiredo\}@dcc.ufmg.br, pjamshid@cse.sc.edu,   kaestner@cs.cmu.edu}}

\maketitle

\begin{abstract}
% Software developed in different platforms has different characteristics and needs. More specifically, code changes are differently performed in the mobile platform compared to non-mobile platforms (e.g., desktop and Web platforms). Prior works have investigated the differences in specific platforms. However, we still lack a deeper understanding of how code changes evolve across different software platforms. In this paper, we present a study aiming at investigating the frequency of changes and how source code changes, build changes and test changes co-evolve in mobile and non-mobile platforms. We developed linear regression models to explain which factors influence the frequency of changes in different platforms and applied the Apriori algorithm to find types of changes that frequently occur together. Our findings show that non-mobile repositories have a higher number of commits per month compared to mobile and our regression models suggest that being mobile significantly impacts on the number of commits in a negative direction when controlling for confound factors, such as code size. We also found that developers do not usually change source code files together with build files or test files. We argue that our results can provide valuable information for developers on how changes are performed in different platforms so that practices adopted in successful software systems can be followed.

Code changes are performed differently in the mobile and non-mobile platforms. Prior work has investigated the differences in specific platforms. However, we still lack a deeper understanding of how code changes evolve across different software platforms. In this paper, we present a study aiming at investigating the frequency of changes and how source code, build and test changes co-evolve in mobile and non-mobile platforms. We developed regression models to explain which factors influence the frequency of changes and applied the Apriori algorithm to find types of changes that frequently co-occur. Our findings show that non-mobile repositories have a higher number of commits per month and our regression models suggest that being mobile significantly impacts on the number of commits in a negative direction when controlling for confound factors, such as code size. We also found that developers do not usually change source code files together with build or test files. We argue that our results can provide valuable information for developers on how changes are performed in different platforms so that practices adopted in successful software systems can be followed.
\end{abstract}

\begin{IEEEkeywords}
Software evolution, code changes, platforms.
\end{IEEEkeywords}

\input{1-introduction.tex}
\input{3-method.tex}

\input{4-results.tex}
\input{7-relatedWork.tex}
\input{8-finalRemarks.tex}

\footnotesize{}
\bibliographystyle{IEEEtranN}
\bibliography{0-main}

\end{document}

%% file: 1-introduction.tex
\section{Introduction}
\label{intro}

% Context and brief definitions and considerations
Software systems developed in different platforms have different practices due to their specific needs~\citep{murphy2014cowboys}. Software platform (e.g., mobile and desktop) refers to the underlying "structure" upon which software is built and it has specific characteristics~\citep{zhang2018within}. For instance, the mobile platform, differently from desktop and Web applications, is usually used to develop sensor-, gesture-, and event-driven applications and it has memory and power consumption constraints~\citep{zhang2018within}. Also, in desktop, the most frequent high-severity bugs occur due to build issues, while in Android, the cause of most problematic bugs is concurrency~\citep{zhou2015cross}.

% Different software platforms present different characteristics. For example, bug causes and bug fixing processes are different in desktop and Android. While in desktop, the most frequent high-severity bugs occur due to build issues, in Android, the cause of most problematic bugs is concurrency~\citep{zhou2015cross}. 
%In this paper, we analyze the commit history of GitHub projects to investigate how software in different platforms is maintained and evolved.

% Works so far
% The diversity in software engineering has motivated several works from the research community and also industry-track works that investigated different practices in specific software platforms~\cite{zhou2015cross,murphy2014cowboys,richardson2016healthcare,russo2017banking,segura2014automated,wright2012release}. However, to the best of our knowledge, we still lack a more comprehensive understanding of how code changes evolve across different software platforms.

% Problem and motivation
%Considering software engineering as a homogeneous whole may be problematic as the needs of systems from specific platforms are ignored~\citep{murphy2014cowboys}. 
Ignoring differences from software platforms may be problematic. Shedding light on software evolution of specific platforms may improve the development of platform-specific tools. It may also strongly benefit developers as they can be aware of how that platform works and behaves before getting into it, mainly those professionals who are looking for a new job. For instance, if a professional is looking for a mobile developer position, a previous understanding of common patterns of changes (e.g., types of code changes usually made together) may prepare the developer for an interview.

Revealing how code changes are performed may also support newcomers who intend to contribute to Open Source Software (OSS) projects, given their important role in the survival and long-term success of community-based OSS~\citep{steinmacher2016overcoming}. Due to the quite independent and self-organized characteristics of working in OSS projects~\citep{steinmacher2016overcoming}, newcomers should be provided with insights and technical support of how current contributors in fact work so that they can be prepared. Prior work has addressed this issue~\citep{steinmacher2016overcoming}, but it has not identified significant ways to support newcomers with technical barriers.

For instance, understanding how changes are performed in a repository before sending pull requests or joining an OSS project may provide the newcomer with valuable information to support this initial phase of contribution and avoid rework or contribution rejection, which could demotivate the developer to keep contributing. We argue that providing insights at platform level is an additional useful information to the newcomer, besides other library and framework-related information.
%As a concrete example, we can imagine that if developers from a certain open source repository usually change source code files together with configuration files, a developer who intends to submit a source code change via pull request should also provide a change in configuration files.

Another important aspect of understanding changes is concerning the specific behavior of the development team depending on the platform. Few works have studied the differences in desktop and mobile platforms~\citep{zhang2018within,bhattacharya2013empirical}. Desktop system developers usually are not involved in reporting bugs and the bug-fixing process takes a longer period of time compared to Android and iOS~\citep{zhang2018within,bhattacharya2013empirical,breu2010information}. On the other hand, bug fixers of mobile applications are more involved in reporting bugs to be discussed and the main causes of bugs are concurrency (in Android) and application logic (in iOS)~\citep{zhang2018within}. We believe companies should provide targeted training for their employees to focus on specific platform' characteristics and needs, and on how developers from those platforms usually work (behavior and practices adopted).

% What we do... In this paper....
This paper presents a study aiming at understanding how code changes evolve in mobile and non-mobile platforms. At this stage, we focus on Android projects (mobile) and Java-based desktop/Web projects (non-mobile). We hypothesize the mobile platform has different evolution patterns compared to non-mobile platforms. The analyses are performed on a dataset composed of 363 popular OSS systems from GitHub: 181 Android applications and 182 desktop and Web applications. We investigate the frequency of commits, whether being mobile significantly impacts on the frequency, and the co-evolution of three different sorts of changes: source code changes, build changes, and test changes. \\

% findings
Our findings, while preliminary, show that non-mobile repositories have a higher number of commits per month compared to mobile. The trend graphs for both platforms are not similar, but both have a peculiar behavior in the holiday season (period from November to January): the number of commits sharply decreases. Our regression models suggest that being mobile significantly impacts on the number of commits in a negative direction when controlling for confound factors.
%The Cohen's $f^2$ measure is 0.19, indicating a medium effect size of the mobile variable in our models.

% Section~\ref{method} presents the methodological procedures we follow in this study. In Sections~\ref{results} and ~\ref{discussion}, we present and discuss our results. Section~\ref{threats} discusses threats to validity and Section~\ref{relatedWork} discusses prior related works. Finally, we conclude this paper in Section~\ref{final_remarks}.

%% file: 3-method.tex
\section{Methodological Procedures}
\label{method}

Our goal in this study is to understand how code changes evolve in mobile and non-mobile platforms. Regarding the mobile platform, we consider only Android applications since they are largely present in GitHub and we were able to find several repositories. For the non-mobile platform, we consider desktop and Web applications as we intend to compare mobile platform against other platforms. That is, we do not aim at comparing all platforms against each other. We address the following initial research questions:

\begin{itemize}
\item \emph{RQ1: How frequent are code changes in mobile and non-mobile platforms?}

%\item \emph{RQ2: How scattered and deep are code changes in mobile and non mobile applications?}

\item \emph{RQ2: How is the co-evolution of source code changes, build changes and test changes in mobile and non-mobile platforms?}
 
\end{itemize}

\subsection{Study Phases}
\label{phases}
To answer RQ1, we rely on statistical modelling (linear regression) to understand the frequency of commit activity in repositories from mobile and non-mobile platforms. To address RQ2, we make use of \textit{Apriori} to analyze the co-evolution of code changes made to three types of files: source code, build, and test files. The study is composed of three main phases, detailed next.

\noindent\textbf{Phase 1 - Software Repository Mining. }We initially selected the 1000 most popular Java repositories in GitHub based on their number of stars, which is considered a reliable proxy to the repository popularity~\citep{borges2016understanding}. 
%We focus on Java systems due to constraints in our data analysis phase. For instance, we analyze the evolution of changes in build files, including files from the Apache Maven, which is a build automation used primarily for Java projects. %Repositories were retrieved from GitHub between July and August 2018.
%The set of 1000 repositories includes different types of systems, such as tools, libraries and software systems designed for the Android platform.
Aiming at retrieving the most relevant repositories, we filtered out repositories with less than 1000 SLOC (toy samples) and we consider only repositories with at least 24 commits in the last 2 years (active projects). This filtering process resulted in 363 repositories. We automatically classified these systems as mobile (if it contains \textit{AndroidManifest.xml})  or non-mobile. Our final dataset contains 181 mobile systems and 182 non-mobile systems. Table I presents the aggregate statistics of our dataset. We can see the number of stars, SLOC, number of contributors, number of pull requests, and number of issues. The systems in our dataset are relevant as indicated the mean number of starts (i.e., more than 6,000 for both platforms). Furthermore, regarding SLOC, we can see that systems in non-mobile platforms are larger than systems in mobile, with means of 152K SLOC and 40K SLOC, respectively.

% \usepackage{multirow}
\begin{table}[h]
\label{dataset_stat}
\centering
\caption{Aggregate statistics of the 363 repositories}
\begin{tabular}{p{1.2cm}p{1cm}p{0.9cm}p{0.9cm}p{0.5cm}p{0.7cm}p{0.7cm}}\Xhline{3.5\arrayrulewidth}
                                      &                        & \textbf{Mean} & \textbf{St. Dev.} & \textbf{Min} & \textbf{Median} & \textbf{Max} \\\Xhline{3.5\arrayrulewidth}
\multirow{5}{*}{\textbf{Mobile}}      & \textbf{Stars}         & 6308.32      & 4573.52          & 2451         & 4710            & 24975        \\
                                      & \textbf{SLOC}          & 40706.21      & 191940.8          & 1003         & 7807            & 2367689      \\
                                      & \textbf{Contribut.}  & 43.73      & 64.65          & 1            & 21              & 351          \\
                                      & \textbf{Pull Req.} & 9.54      & 16.69          & 0            & 3               & 84           \\
                                      & \textbf{Issues}        & 125.98      & 193.72          & 0            & 65              & 1640         \\\Xhline{3.5\arrayrulewidth}
\multirow{5}{*}{\textbf{Non-Mobile}} & \textbf{Stars}         & 6490.28       & 6426.73          & 2443         & 4548            & 41653        \\
                                      & \textbf{SLOC}          & 152319.4      & 295851.9          & 1418         & 48158.5         & 2729887      \\
                                      & \textbf{Contribut.}  & 96.69      & 94.05          & 1            & 64              & 400          \\
                                      & \textbf{Pull Req.} & 30.06      & 62.73          & 0            & 9               & 521          \\
                                      & \textbf{Issues}        & 231.83      & 304.37          & 0            & 120             & 1730        
\end{tabular}
\end{table}

\noindent\textbf{Phase 2 - Data Collection. } Through the GitHub REST API\footnote{https://developer.github.com/v3/}, we collected the following data at repository-level: number of contributors, number of pull requests, number of issues, and SLOC. Furthermore, we collected commit-level data by the mining algorithm to perform the code change co-evolution analysis: commit date and type(s) of file(s) changed by the commit (source code, build or/and test files). In addition, we retrieved additional information (number of changed files, added lines of code, deleted lines of code, and total changed lines of code) to be used in next steps of our work.

\noindent\textbf{Phase 3 - Data Analysis. }We have two main parts in the data analysis. First, we use statistical modelling to address the first research question. Second, we apply the \textit{Apriori} algorithm to check whether different types of code changes co-occur in commits. Next, we detail how we proceed when building linear regression models and applying the mining algorithm.

%%%%%%%%%%%%%%%%%%%%%%%%%%%%%%%%%%%%%%%%%%%%%%%%%%%%%%%%%%%%%%%%%%%%%%%%%%%%%%%%%%%%%%%%%%%%%%%%

\subsection{Statistical Modelling}
\label{statistical}

% To provide evidence on whether being mobile influences the frequency of commits, we developed two multiple linear regression models, one with only the control variables and a full model with the indicator and control variables. By controlling for confound factors in the multiple regression, we evaluate whether the difference in the frequency of commits can be attributed to the fact of a repository being mobile or not. Our hypothesis is that mobile systems have a higher frequency of commits since users from that platform (in our case, Android users) expect fast bug fixes and rapid availability of new features~\citep{oliveira2018empirical,banerjee2016automated}.

Our hypothesis is that mobile systems have a higher frequency of commits since users from that platform (in our case, Android users) expect fast bug fixes and rapid availability of new features~\citep{oliveira2018empirical,banerjee2016automated}. To provide evidence on whether being mobile influences the frequency of commits, we build two successive regression models: a model that contains only the control variables; and a model with the addition of the experimental (indicator) variable. We then use Cohen's $f^2$ measure to gauge the effect size of the indicator variable. We consider model coefficients important if they are statistically significant at a 0.05 level. In our models, the response (dependent) variable is the number of commits per month - \textbf{nCommMonth}. Based on variables that can influence the number of commits, we consider the following repository-level control (independent) variables: size of the system in terms of number of source lines of code - \textbf{sloc}, number of contributors - \textbf{nCont}, number of pull requests - \textbf{nPR}, and number of issues - \textbf{nIssues}. We also have an indicator (experimental) variable, \textbf{isMobile}, which is a binary variable that indicates whether a repository is mobile (1) or not (0).

Before building our models, we log-transformed variables aiming at stabilizing their variance and reduce heteroscedasticity~\citep{zhang2018within,cohen2014applied}. To compare the distribution of our raw data regarding number of commits per month for both groups (mobile and non-mobile), we adopt the non-parametric Wilcoxon Signed-Rank Test. We also report the Cliff's delta to indicate the size of the difference of distributions. Finally, to tackle possible problems related to multicollinearity~\citep{farrar1967multicollinearity} in our regression analysis, we check for multicollinearity using the variance inflation factor (VIF)~\citep{allison1999multiple}. If it is below 3, which is a safe and conservative threshold, we confirm that our models do not suffer from multicollinearity~\cite{zhang2018within,trockman2018adding}.

% To compare the distribution of our raw data regarding number of commits per month for both groups (mobile and non-mobile), we adopt the non-parametric Wilcoxon Signed-Rank Test. We also report the Cliff's delta to indicate the size of the difference of distributions. Mobile systems may (apparently) impact on the number of commits per month, but underlying confound factors might actually be leading the response of our models. We build two successive regression models: a model that contains only the control variables; and a model with the addition of the experimental (indicator) variable. We then use Cohen's $f^2$ measure to gauge the effect size of the indicator variable. We consider model coefficients important if they are statistically significant at a 0.05 level. Finally, to tackle possible problems related to multicollinearity~\citep{farrar1967multicollinearity} in our regression analysis, we check for multicollinearity using the variance inflation factor (VIF)~\citep{allison1999multiple}. If it is below 3, which is a safe and conservative threshold, we confirm that our models do not suffer from multicollinearity~\cite{zhang2018within,trockman2018adding}.

%%%%%%%%%%%%%%%%%%%%%%%%%%%%%%%%%%%%%%%%%%%%%%%%%%%%%%%%%%%%%%%%%%%%%%%%%%%%%%%%%%%%%%%

\subsection{Mining frequent code changes and association rules}
\label{mining_method}

To find co-occurrences of different code changes file types along the last 2 years, we apply Apriori~\citep{agrawal1996fast,agrawal1993mining}. 
% As this mining algorithm is order-insensitive~\citep{molderez2017mining}, we believe it is suitable for our study since we do not need ordered data. 
We analyze whether there are co-occurrences of source code changes, build changes, and test changes. After obtaining the co-occurrences of code changes, we are able to find association rules also using the \textit{Apriori} algorithm. Therefore, based on a change the developer performs, we can suggest other types of changes according to the learned association rules. 
% Using \textit{Apriori} algorithm requires the specification of a support value, as explained in Section~\ref{frequent_mining}.

We rely on heuristics to identify build and test changes. For build changes, we look for files with names as recommended by the build automation tools we use. For Apache Maven build files, we search for \textit{pom.xml}; for Apache Ant files, we look for \textit{build.xml}; finally, for Graddle files, we search for \textit{build.graddle}. Regarding changes on test files, we adapt an heuristic adopted by previous works~\citep{zaidman2011studying,levin2017co}. We classify a change as a test change if the name of the class begins with the word "Test" or ends with the word "Test", or "Tests", or "TestCase". We also consider a test change if the modified class is contained in a directory with the word "Test", "Tests", or "TestCase". Note that all situations in which the word is lower case are considered in the same way.

%% file: 4-results.tex
\section{Results and Discussion}
\label{results}

\subsection{Frequency of Commits}
\label{results_frequency}
\begin{figure}[!h]
\centering
\includegraphics[width=0.45\textwidth]{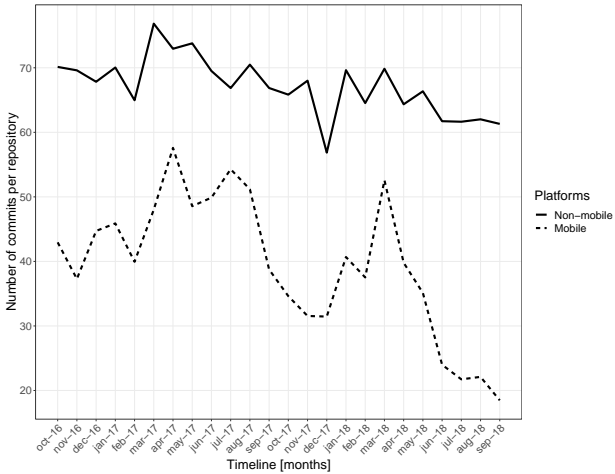}
\caption{Frequency of commits in a 2-year time period.}
\label{fig:plot_frequency}
\end{figure}

We analyzed 465,500 commits from 363 repositories hosted in GitHub. Figure~\ref{fig:plot_frequency} presents the frequency of commits (average number of commits per month per repository) in mobile and non-mobile platforms in 2 years. We double checked the data to confirm the sharp drop in the end of the plot, which may be caused by discontinuation of mobile applications.

% \begin{figure}[!h]
% \centering
% \includegraphics[width=8cm,height=6cm,keepaspectratio]{img/plot_frequency.pdf}
% \caption{Frequency of commits in a 2-year time period.}
% \label{fig:plot_frequency}
% \end{figure}

Surprisingly, the number of commits is always higher in non-mobile platform compared to mobile. For the mobile platform, we can note a regular pattern in some periods, specially in the holiday season (November to January). For instance, in the period from nov-16 to apr-17, the number of commits increased about 54\%. The curve behaves similarly in the period from nov-17 to mar-18, with an increase of approximately 67\%. This may suggest that some factors influence this behavior and contributions to OSS mobile projects in that period of the year. Regarding those periods in non-mobile platforms, the increase in the number of commits was much smaller, with 4.8\% for nov-16 to apr-17 and 2.7\% for nov-17 to mar-18. However, we can observe that non-mobile platforms had a very low average number of commits in December 2017 (56 commits), which suggests that holiday season may also influence the work activity in non-mobile projects. 

This kind of temporal picture helps us to see the general trends, and how both platforms behave along these 2 years. However, we still lack an explanation regarding what factors are impacting the frequency. We then developed multiple linear regression models to understand the impact of the platforms on the frequency when controlling for confound variables.

\begin{figure}[!h]
\centering
\includegraphics[width=0.4\textwidth]{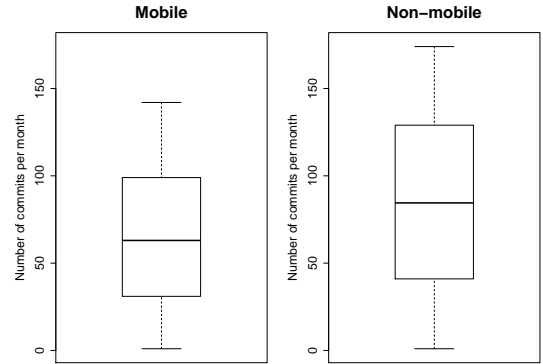}
\caption{Distributions of response variable for mobile and non-mobile.}
\label{fig:boxplot_distributions_frequency}
\end{figure}

\noindent\textbf{Distribution comparison. }Figure~\ref{fig:boxplot_distributions_frequency} presents the boxplots corresponding to the distribution of commits per month (response variable) for both platforms: mobile and non-mobile. We can see that the two distributions are different. In fact, the median number of commits per month for mobile is approximately 63, while for non-mobile is 84. In addition, we obtained a Cliff's Delta of -0.2181 (small), with a 95\% confidence interval, indicating a small but statistically significant difference.

% \begin{figure}[!h]
% \centering
% \includegraphics[width=0.4\textwidth]{img/boxplots.pdf}
% \caption{Distributions of response variable for mobile and non-mobile.}
% \label{fig:boxplot_distributions_frequency}
% \end{figure}

\noindent\textbf{Additional explanatory power.}
Table~\ref{tab:regression_coeff} presents our model coefficients along with their p-values. From the model with only control variables, we observe that most coefficients are in the positive direction (positive \textit{T value}), as expected. For instance, we expect that more contributors result in more commits per month. The same is valid for number of pull requests and number of issues. The unexpected results occur for \textit{sloc} coefficient, as its signal is negative. By inspecting the significance, apart from the intercept coefficient, all coefficients are not significant. We checked for correlation of control variables with the response variable and in fact they are not highly correlated. The highest correlation value occurs for number of contributors and number of commits per month (pearson coefficient of 0.1992).

\vspace{-0.4cm}
% \usepackage{multirow}
\begin{table}[h]
\caption{Multiple linear regression coefficients for our two models.}
\centering
\begin{tabular}{clll}\Xhline{3.5\arrayrulewidth}
\multicolumn{1}{l}{}                                   
& \textbf{Variable}   & \textbf{T value}   & \textbf{P value (significance)}\\\Xhline{3.5\arrayrulewidth}

& (Intercept)     & 61.02      & \textless 2e-16 ***    \\
& nCont     & 1.283    & 0.2 \\
& sloc      & -0.224     & 0.823   \\
& nPR      & 0.235     & 0.814   \\

\multirow{-5}{*}{\textbf{\begin{tabular}[c]{@{}c@{}}Control variables\\ only\end{tabular}}}                     
& nIssues     & 1.233       & 0.218     \\\Xhline{3.5\arrayrulewidth}

& (Intercept)        & 66.194      & \textless 2e-16 *** \\
 
%& {\color[HTML]{FE0000} Indicator variable} & {\color[HTML]{FE0000} -15.246} & {\color[HTML]{FE0000} \textless{}2e-16 ***} \\
& isMobile & -15.246 &  \textless{}2e-16 *** \\

& nCont      & -1.19        & 0.235   \\
& sloc      & -1.525    & 0.128      \\
& nPR         & -0.117       & 0.907  \\

\multirow{-6}{*}{\textbf{\begin{tabular}[c]{@{}c@{}}Full model,\\ including indicator\\ variable\end{tabular}}} 
& nIssues       & 0.751   & 0.453    \\\Xhline{3.5\arrayrulewidth}

\multicolumn{1}{l}{}    & \multicolumn{3}{l}{***p \textless 0.001, **p \textless 0.01, *p \textless 0.05}  
\label{tab:regression_coeff}
\end{tabular}
\end{table}

To gauge the effect of the indicator variable (\textit{isMobile}), we build a successive regression model including the binary variable. In Table~\ref{tab:regression_coeff}, we can also see the coefficients of the full model. In fact, the indicator variable has a statistically significant impact on the frequency of commits in repositories (p-value \textless 2e-16). The indicator variable also increased the explanatory power of the model, as suggested by a proportional change in $R^2$ of 1,900\% (from 0.02 to 0.4). We adopted the Cohen's $f^2$ measure to estimate the effect size of the indicator variable. We computed the $R^2$ for both models (controls and indicator, and only controls) and obtained a Cohen's $f^2$ of \textbf{0.19}. The following thresholds are suggested to indicate the effect size: 0.02 (small), 0.15 (medium), and 0.35 (large). We can therefore conclude that the effect of being a mobile repository on the frequency of commits when controlling for confound variables is \textbf{medium}.

\noindent\textbf{Multicollinearity diagnosis. } We diagnose our models, checking for multicollinearity, since highly correlated regressors may inflate the variance. We first check the correlation between the predictors and then we get the variance inflation factor (VIF) for each predictor. 
% Figure~\ref{fig:corr_matrix} presents a matrix-style image with the correlation between all pairs of predictors in our models. 
We noticed that predictors are not highly correlated (the highest correlation is 0.6 between number of pull requests and number of issues). Regarding the variance inflation factor, all variables have VIF values below 3, which is a safe value and indicate that our models do not suffer from multicollinearity~\cite{zhang2018within,trockman2018adding}.

% \begin{figure}[!h]
% \centering
% \includegraphics[width=0.5\textwidth]{img/correlation_matrix.pdf}
% \caption{Correlations between predictor variables.}
% \label{fig:corr_matrix}
% \end{figure}

% Eduardo says: I think this paragraph can be (shorten and) moved to the conclusion. It just repeats the key results of this section.
% We observed code changes are more frequent in non-mobile platforms when compared to mobile. By analyzing the trend graph over the past two years, we note that non-mobile repositories have more commits per month in all 24 months compared to mobile repositories. Furthermore, we can observe a regular pattern of change in the mobile platform, with a possible seasonal behavior. Our multiple linear regression models indicate that being mobile significantly impacts the frequency of commits when controlling for the following confound variables: number of contributors, size of system (in SLOC), number of pull requests, and number of issues.

% \begin{framed}
% \noindent\textbf{Answering RQ1: }Code changes are more frequent in non-mobile platforms compared to the mobile platform. Furthermore, being mobile significantly impacts (in the negative direction) the frequency of commits when controlling for confound variables.
% \end{framed}

\subsection{Frequent Code Change Types and Association Rules}
\label{results_frequentChanges}
Regarding the types of frequent code changes, we analyzed the commit history using the \textit{Apriori} algorithm and set a minimum support value of 0.05. 
%Our minimum support must be a low value given the characteristics of our dataset, in which commits are much more likely to change a single type of file. More specifically, commits usually change only source code files (67\% of mobile changes are source code changes and 70\% of non-mobile changes are source code changes). Given these characteristics, we may expect that \textit{support} metric values are low.
When analyzing the association rules, we focus on the \textit{confidence} and \textit{lift} metrics to check the strength of the rules. Table~\ref{tab:frequent_changes} presents the code changes that co-occur along with their support and absolute count values. 
%As we can see, each type of change appeared individually as a frequent type, but we are interested in co-occurrences of different types of changes.

% \usepackage{multirow}
\begin{table}[t]
\caption{Frequent types code changes in all commits.}
\centering
\resizebox{0.5\textwidth}{!}{%
\begin{tabular}{clll}\Xhline{3.5\arrayrulewidth}
\multicolumn{1}{l}{}                 & \textbf{Frequent change types} & \textbf{Support} & \textbf{Absolute count} \\\Xhline{3.5\arrayrulewidth}
\multirow{4}{*}{\textbf{Mobile}}     & \{build\}                      & 0.07081202       & 12166                   \\
                                     & \{test\}                       & 0.07848341       & 13484                   \\
                                     & \{source\_code\}               & 0.67038014       & 115176                  \\
                                     & \{source\_code,test\}          & 0.05435169       & 9338                    \\\Xhline{3.5\arrayrulewidth}
\multirow{5}{*}{\textbf{Non-mobile}} & \{test\}                       & 0.09728867       & 28573                   \\
                                     & \{build\}                      & 0.11002986       & 32315                   \\
                                     & \{source\_code\}               & 0.70661541       & 207528                  \\
                                     & \{source\_code,test\}          & 0.05106353       & 14997                   \\
                                     & \{source\_code,build\}         & 0.05217012       & 15322                  
\end{tabular}}
\label{tab:frequent_changes}
\end{table}

We can note that all types of changes, when considered individually, appear in the results returned by the algorithm. However, we focus only on types of changes that occur together with other types. That is, we analyze results where at least two types of changes appear. In mobile systems, the types of code changes that occur together (i.e., in the same commit) with a support greater than the minimum support is source code and test changes. This co-occurrence happened with a support of 0.054 (i.e., in 5.4\% of all commits). The low support indicates that developers do not usually perform changes in source code and test files simultaneously. Surprisingly, mobile developers do not usually change source code files together with build files~\citep{macho2017extracting}. We performed some tests and found that source code changes occur together with build changes only with a support of 0.03. Regarding non-mobile platforms, we see two co-occurrences of types of changes. First, source code changes occur together with test changes with a support of 0.051. Second, source code changes also co-occur with build changes with a similar support: 0.052. 

To obtained association rules, we rely on default values of minimum support (0.001) and minimum confidence (0.8) defined by the \texttt{arules} package in R. We then found the following rule for the mobile platform: \textit{\{build,test\} => \{source code\}}. This association rule has a support value of 0.0062, confidence of 0.9177, lift of 1.3689, and absolute count of 1070. This rule indicates that developers commonly perform changes in source code files given they changed build and test files. However, the low value of support shows that both sides of the rule (build and test changes, and source code changes) do not occur very frequently in commits. 

A similar rule was obtained for non-mobile platforms: \textit{\{build,test\} => \{source code\}}. This association rule has a support value of 0.0134, confidence of 0.8597, lift of 1.2166, and absolute count of 3927. Although the association rule for non-mobile is the same of mobile, we obtained different metric values. For instance, the support is 2.15 times higher in non-mobile than in mobile, which indicates that both sides of the rules (build and test changes, and source code changes) occur more frequently in commits of non-mobile applications. However, the confidence is lower. This indicates that, although a source code change is likely to be necessary given that build and test changes were made, the strength of this statement is  smaller compared to mobile platform.

As we observed, there is a regular pattern of change in the mobile platform, with a possible seasonal behavior. As a possible \textbf{implication for newcomers} who intend to contribute to mobile repositories,  we recommend that they look at the temporal trend before sending a contribution to an OSS project. For instance, it is not recommended to send a pull request in the end of the year as projects are not very active at that time. Instead, we recommend to send a pull requests in the beginning of the year. We also highlight that source code contributions to non-mobile repositories may require changes to test files or build configuration files as well since changes to such kinds of files usually occur together.

%% file: 7-relatedWork.tex
\section{Related Work}
\label{relatedWork}

Several works have investigated different types of code changes and performed commit history analysis. For instance, ~\citet{levin2017co} analyzed the co-evolution of only test changes and source code changes. In this work, we also included build changes in our co-evolution analysis. In addition,~\citet{macho2017extracting} performed a study on build changes relying only on the Apache Maven build automation tool (\textit{pom.xml} files). On the other hand, we included two other build automation tools: Ant (\textit{build.xml}) and Graddle (\textit{build.graddle}). Furthermore, we compare the co-evolution of changes in mobile applications against non-mobile applications.~\citet{farago2015cumulative} investigated whether modifications performed on frequently changing code have worse effect on software maintainability than those affecting less frequently modified code. Their findings indicated that modifying high-churn code is more likely to decrease the overall maintainability of a software system, which can increase the number of defects.

In summary, prior works focused on different aspects of code changes. Here, we perform a broader analysis of code changes, investigating their frequency, factors that explain them, and the co-evolution of different types of changes.

%% file: 8-finalRemarks.tex
\section{Final Remarks and Future Work}
\label{final_remarks}

This paper presented a study aiming at investigating how code changes evolve in mobile and non-mobile platforms by analyzing the frequency of commits. Our statistical analysis revealed that being mobile significantly impacts the frequency of commits. We also observed that in the mobile platform source code changes occur together with test changes with a low frequency. In non-mobile platforms, we found that (i) source code changes and test changes and (ii) source code changes and build changes. This is an undergoing work, and as a next step, we plan to investigate mobile, desktop, and Web platforms separately. We also intend to perform a time-series analysis to reveal more explicit patterns of evolution of code changes and understand which factors influence the evolution, and extend our study to analyze how scattered (number of changed files) and deep (LOC) code changes are in different platforms. Finally, we plan to interview professionals from each platform.
\vspace{-0.2cm}

\section{ACKNOWLEDGMENTS}

This research work was partially supported by CAPES, CNPq (grant 424340/2016-0), and FAPEMIG (grant PPM-00651-17). K\"{a}stner\textquotesingle s work has been supported in part by the NSF (awards 1318808, 1552944, and 1717022) and AFRL and DARPA (FA8750-16-2-0042).